# Towards a More Systematic Approach to Secure Systems Design and Analysis

Simon Miller, University of Nottingham, UK.
Susan Appleby, CESG, Cheltenham, UK.
Jonathan M. Garibaldi, University of Nottingham, UK.
Uwe Aickelin, University of Nottingham, UK.

## Abstract

The task of designing secure software systems is fraught with uncertainty, as data on uncommon attacks is limited, costs are difficult to estimate, and technology and tools are continually changing. Consequently, experts may interpret the security risks posed to a system in different ways, leading to variation in assessment. This paper presents research into measuring the variability in decision making between security professionals, with the ultimate goal of improving the quality of security advice given to software system designers. A set of thirty nine cyber-security experts took part in an exercise in which they independently assessed a realistic system scenario. This study quantifies agreement in the opinions of experts, examines methods of aggregating opinions, and produces an assessment of attacks from ratings of their components. We show that when aggregated, a coherent consensus view of security emerges which can be used to inform decisions made during systems design.

## Introduction

Today, an ever-growing number of sensitive transactions of data take place on-line (e.g., e-government, internet banking and e-commerce), and cyber crime has become prevalent. One of the consequences of this is that the cybersecurity of information systems has become an increasing concern. Assessing the level of risk posed by specific events is an area of ongoing interest for most (if not all) organisations, leading to a requirement for scientific methods of validating the cybersecurity of software systems.

The questions posed by this special issue are:

1. *'What are the foundational, measurable, and repeatable scientific elements applicable to assuring the cybersecurity of software systems?'*

In the real world, the subjective opinions of cybersecurity experts are used to assess the security of software systems in their design stage. Indeed, this is often the only way to make such assessment. However, the human elements involved introduce the potential for inconsistency both between experts, and within the experts themselves. In this paper we show how the opinions of experts can be elicited and measured, and how we can ameliorate (and measure) the effects of their inherent variation through aggregation to produce a consistent and repeatable assessment.

> 2. *'How should we verify and validate software systems in terms that will provide indisputable scientific evidence that they are secure?'*
>
> We contend that 'indisputable evidence' of security is not a practical concept in the real-world, as no system can be guaranteed to be without security risks for any length of time. Furthermore, we argue that independent, measured, 'proven' expert opinion should be used to verify and validate software systems. Repeated successful assessments by cybersecurity experts provide the historical scientific evidence that their opinions are a valid and proven method of assuring the security of software systems. The job of security practitioners is to make an informed judgement as to whether the system is secured to an appropriate degree for the threat environment it faces.

The work described in this paper addresses two key topics of interest, *Measuring Human Factors in Security* – In the method we present, human experts are used to validate software systems. Perceptions of security vary both between experts, and within an individual expert. Our method allows us to explicitly measure this variation, and produce an assessment that accounts for it. *Quantitative Security Management* – The outcome of the proposed method is a quantification of expert opinion of the security of a system, including a measurement of uncertainty. These values can be used to make decisions regarding the implementation of a software system, and the management of its security framework.

In the proposed method we use two different types of survey to elicit the opinions of a group of security experts. The first involves ranking a series of technical attacks on a system in order of how difficult they are to carry out undetected by the system or its operators. The second requires experts to rate components of attacks in terms of aspects which are thought to contribute to their overall difficulty. In practice, a system is only as secure as its weakest element, i.e., the easiest way in. Identifying which are the weakest aspects of a system, i.e., the 'easiest' ways of attacking it, is thus a highly relevant component of system security assessment, though obviously it does not provide all of the answers.

We have applied the method to measure variation within a set of thirty nine highly experienced expert practitioners including system and software architects, security consultants, penetration testers, vulnerability researchers and specialist systems evaluators.

This paper shows how we are able to use expert opinion to produce a consistent assessment, and identify outliers. We also discuss the meaning of the results, and how the approach could be applied in future work.

## Motivation

Designing and assessing a secure systems architecture is an increasingly complex problem due to the diversification and expansion of technology which has taken place over the last few years. For example, in recent years there has been a vast expansion in use of shared/collaborative services, and in the use of virtualisation technologies such as in provision of cloud services. Architectures are more fluid and there are few established security models.

Defensive security technologies have also become more complex, and in some cases it is far from clear what a product actually does or how well it does its job. Trends which have complicated the picture include the greater use of after-the-fact heuristic techniques in products such as anti-virus or IDS systems, the greater use of isolation techniques such as sandboxing, and complex aggregation and analysis of observed data by security vendors.

Threat assessment has also become more complex, as sophisticated attack techniques have proliferated via toolsets available on the Internet, placing sophisticated capability into unsophisticated hands. These rapid changes make it difficult for experts to make consistent, well founded judgements about what secure architecture and design practices to adopt.

## Background

While others have examined the aggregation of experts' opinions for security assessment, we are not aware of any work that assesses the variation in security expert decision making with regard to ranking attacks by difficulty, and rating how difficult it is to compromise/bypass their components. Related work detailing the aggregation of experts' opinions for security assessments will be described, as this task shares many similarities with the problem addressed by the proposed method.

Tan and Li (2012) propose a combination of Analytical Hierarchy Process (AHP) and information entropy for representing group decision making when assessing security risks. AHP (Saaty, 1990) is a method of decision making that involves breaking problems down into a hierarchy of more manageable problems, which are then compared to assign a relative weighting and assessed individually. Then, a level of risk for the overall problem is calculated using the individual assessments and weightings. In this case experts' weightings are calculated using information entropy, which takes into account their professional status and the credibility of their submitted opinion. Their opinions are aggregated using a process involving a weighted geometric mean that produces an overall opinion. The authors provide an example showing how three experts' opinions of the level of risk posed by a set of security threats are combined using the proposed method.

A method proposed by Chan (2010) uses a Bayesian index to combine experts' opinions of security risk into one information security (IS) risk model. Bayesian models provide the means to compute the probability of high or low information risk based on a set of risk indicators. In this study, eleven experts created a list of security risk indicators and assigned weights to them. The resulting Bayesian model was validated with forty one companies, each of which completed a survey regarding their IS protection measures, and the occurrence of IS incidents during the past two years. The results show that there was high correlation between the companies' Bayesian indexes and experts' judgements.

Goyette and Karmouch (2011) propose a method of assessing the security of virtual networks. This is a particularly problematic area of information security as service providers do not have details of the physical infrastructure over which the virtual network operates. A combination of Dempster-Shafer theory (Dempster, 1968) and MACBETH (Measuring Attractiveness by a Categorical Based Evaluation Technique) are used to provide the means for experts to make asynchronous contributions to an IS model. Dempster-Shafer theory is a method of combining evidence from multiple sources to produce a degree of belief about a question, and MACBETH is a decision making tool used by a group of experts to rank alternatives that depend upon multiple criteria. Experts submit their opinions through a series of questions and a confidence index in their answers. The experts' opinions are aggregated using Dempster-Shafer theory to produce a set of degrees of belief. The authors demonstrate the model with a numerical example involving five experts, where their opinions on one value judgement are fused in a way that does not require the experts to be in the same physical location, or indeed submit their opinions at the same time.

Secure systems design is a problem with inherently high levels of uncertainty. For example, the potential vulnerabilities of products, and how they may evolve over time, must be estimated. Potential new attacks, including previously unseen categories of attacks must also be estimated – and this may be over long timescales if considering the service life of the system. Fuzzy Logic (Zadeh, 1973) is particularly well suited to tasks of this nature as it allows us to model the uncertainty that is present in information systems security problems and the experts who make decisions regarding them. A variety of methods based on fuzzy techniques have been applied in areas closely related to that being addressed in this paper.

A method of risk analysis that uses similarity measures with fuzzy sets is proposed by Chen & Chen (2003). Security risks of a system component are rated by looking at the risk of failure of each of its sub-components, and the severity of losing them. Experts rate each sub-component using linguistic terms (represented by fuzzy sets), and overall risk for a component is computed using a weighted average of the risks associated with the sub-components. The result of the averaging process (a fuzzy set) is then compared to nine linguistic terms using a similarity measure. The most similar term is selected to describe the risk for the component. An example shows how multiple experts' opinions along with their degree of confidence in their assessments can be used to produce an overall risk assessment for a component. The method is shown to be a suitable method of aggregating risks associated with sub-components by multiple experts to form a group opinion of an entire system.

In Garibaldi and Ozen (2007) the authors propose the use of 'nonstationary' fuzzy reasoning to model the variation in expert decision making in a medical case study. Nonstationary fuzzy systems introduce small variations to the model over time, mimicking the temporal variation found in the opinions of real-world experts. This work builds on work in which a standard fuzzy system is used to model medical experts' opinion in the context of umbilical acid base analysis, whereby properties of blood taken from the umbilical cord after child birth are used to determine the health of a new born child. A set of rules for the system was created through consultation with experts, and the system was tuned using the experts' judgements. Fifty cases that were said to be difficult to interpret were selected from a database of ten thousand cases, and each expert was asked to rank them in order from 'worst' to 'best' in terms of health. The standard fuzzy expert system was then used to produce scores for each of the fifty cases, allowing a ranking to be produced for comparison with the experts' rankings. The results show that the ranking produced by the fuzzy expert system was very close to that produced by the experts. Following this, the authors show that by introducing variation into the model, the variation within a single expert over time, and between a group of experts, can be modelled with an approach that does not produce the same answer every time, despite the same input.

Sendi et al. (2010) propose the use of FEMRA (Fuzzy Expert Model for Risk Assessment), a system that represents the knowledge of experts, and uses a fuzzy rule base to produce a numeric value representing risk. A list of assets and threats were identified and three experts were asked to rate the qualities 'Confidentiality', 'Integrity' and 'Availability' in the range [0,1]. A list of vulnerabilities was then created, and used with the list of assets and threats to create a list of risks. Asset values, vulnerability effects and threat effects were rated by the experts, then the fuzzy model took these values and converted each of them into one of three fuzzy sets 'Low', 'Medium' and 'High'. A rule base was created which allowed the combination of these sets to produce an output set determining risk. The final output of the system is an index of risk value produced by 'defuzzifying' the output set, that can be used by managers to decide on the appropriate action to be taken.

Feng and Li (2010) put forward an Information Systems Security (ISS) risk assessment model that uses fuzzy sets to represent the degree of belief for a statement based on current evidence. The model is demonstrated using a real world case study of a Chinese financial service's information systems. Six experts rated the strength of the evidence, in this case components/effects of risks, which are represented using fuzzy sets. For comparison three other approaches were tested: Fuzzy Comprehensive Evaluation (FCE), Bayesian Networks (BNs) and traditional Dempster-Shafer theory. The authors state that their method is an improvement over the other methods tested as it reduces the uncertainty inherent in conflicting evidence provided by multiple experts.

Wu et al. (2009) demonstrate an improved version of AHP based on fuzzy sets to be used for risk assessment of information systems. Two hierarchies are constructed (as in traditional AHP), one representing the probability of an incident (a combination of threats and vulnerabilities), the other representing the impact of an incident (a combination of recovery costs and severity). Experts are asked to rate the factors at the bottom of the hierarchies, their opinions are represented using fuzzy sets. These opinions are then combined to complete the AHP, producing a risk vector. An example is given in which three experts assess the factors identified in a hierarchy, and these opinions are combined to produce a comprehensive fuzzy judgement matrix. The elements of the matrix are then weighted and used to produce a risk vector.

Fuzzy sets are used for information security risk assessment in research carried out by Fu et al. (2008). Three categories (Asset, Threat and Survivability (Vulnerability)) are rated by experts using nine linguistic variables that quantify the risk (e.g., 'very little' loss or 'very high' threat), the experts' opinions are then aggregated using the Delphi method (Linstone and Turoff, 1975). The Delphi method involves gathering experts' opinions in a number of rounds. After each round, an anonymous summary of the experts' opinions and reasoning is shown, and the experts are asked again for their opinion. The idea is that experts may revise their opinions in light of their colleagues' opinions and reasoning. The technique stops when a pre-defined criteria is met (e.g., number of rounds, consensus reached). An overall fuzzy number is produced using all of the categories of risk, which is then defuzzified to produce a risk index. An example shows how the method works, in which the opinions of three experts are used to calculate a description of risk for a synthetic system.

## Methodology

A key part of this research is the elicitation of opinions from a group of cybersecurity experts about how difficult it is to complete attacks and compromise/bypass components for a given system design. The opinions are then used to produce a consistent measurement.

Our partnership with CESG gives us a level of access to such experts that would be difficult to attain otherwise. As the National Technical Authority for Information Security in the UK, CESG has access to a cadre of specialist security architects and other technical security experts, and can draw on both public and private sector expertise.

The methodology was tested in scenarios involving expert analysis of factors relating to attacking a system's architecture. Three knowledge extraction exercises were performed. The following sections describe the initial exercise, a follow up prototype exercise and the main exercise, respectively.

## Initial Exercise

The purpose of the initial exercise was to develop a scenario consisting of a system together with various methods / vulnerabilities / attacks, realistic enough to permit reasoned assessment, while being difficult to assess fully, even by leading experts. A scenario was created by a senior member of CESG technical staff that is designed to be representative of a fairly mainstream government system. The system involves a range of core services and back end office facilities together with remote sites and mobile access. Core systems hold the most sensitive business information, with assets rated in terms of their value at Business Impact Level 3 (BIL3) following the standard UK government scheme (CESG, 2009). This scale rates the impact of an event from BIL0 (no consequences) to BIL6 (catastrophic).

Six technical experts from CESG were presented with the scenario, the creator of the scenario described the system in detail, showing diagrams of various aspects of the system, and answering experts' questions. The group then carried out a mock security review of the architecture, a task highly familiar to them. As a group they were then challenged to identify ten different ways of mounting an end-to-end attack on the system, identifying all individual attack elements involved. The end to end attacks are termed 'attack vectors' (AVs), and individual elements are termed 'hops'. Once a set of ten AVs had been established, the experts were asked to rank them from one to ten in order of how difficult each was to carry out undetected. This was done without the experts communicating with one another to ensure that each individual gave their opinion without outside influence. Even with the experts' in-depth knowledge of information security this was a task with a fair degree of uncertainty, as much of the detail required to precisely assess difficulty was absent. For example, no information was provided about the exact software and hardware being used in the proposed system. Participants were asked to assume that it was of the typical standard that would be used in a government system of this type. All of the experts involved in the exercise regularly work with UK government BIL3 systems, and so are aware of associated security policy and how it is typically applied to such systems in terms of component configuration, frequency of anti-virus updates, etc. Having ranked the ten AVs, the technical experts were then asked to rate each of the hops by difficulty (either 'low', 'medium' or 'high'), and rate their confidence in their answer (as 'low', 'medium' or 'high'), again, this was conducted in isolation. The confidence rating was provided to allow the experts to show uncertainty in their answer, whether it is caused by a lack of clarity in the scenario/hop description, a lack of knowledge of a particular technology (e.g., cryptography), or other sources of uncertainty.

**Data Analysis**

Having collected the data, analysis was performed to examine the variation in opinion within the group. This section contains the outcome of the data analysis activity for this initial exercise. In Table 1, each expert is compared to the group response produced by taking the average ranking of each AV and sorting them into ascending order. The difference column shows the distance between each expert's ranking and the group rank, and Spearman's rho is used to compare each expert's rank order with the group rank order. Finally, Kendall's W is used to compute the rank correlation within the group of all six experts. Spearman's rho measures the statistical dependence of two sets of rankings, correlation is measured on a scale from -1 (perfect negative correlation), through 0 (no correlation), to +1 (perfect positive correlation). Kendall's W is a similar measure used to calculate the agreement between rankings from a group of people, producing values from 0 (no agreement) to 1 (complete agreement). These correlation measures are particularly useful in cases like the one described here, where we are working with a set of subjective rankings, produced by humans.

|                    | A    | B    | C    | D    | E    | F    |
|--------------------|------|------|------|------|------|------|
| **Difference**     | 8    | 14   | 18   | 10   | 6    | 4    |
| **Spearman's Rho** | 0.90 | 0.78 | 0.77 | 0.92 | 0.94 | 0.96 |
| **Kendall's W**    | 0.82 |      |      |      |      |      |

Table 1 - Initial Exercise AV Ranking

The results show that there is a clear positive correlation between the individual rankings and the group rank, and that there is a strong correlation between the individuals themselves. In order to determine whether the averaging processes involved would always result in a high level of correlation, a random set of rankings was produced. For comparison, Table 2 shows the result of the same process using random rankings, which show that the correlation is significantly worse than that produced by the experts.

A Systematic Approach to Secure System Design 8

|               | R1   | R2   | R3   | R4   | R5    | R6   |
|---------------|------|------|------|------|-------|------|
| Difference    | 24   | 16   | 30   | 20   | 42    | 26   |
| Spearman's Rho| 0.25 | 0.75 | 0.30 | 0.61 | -0.15 | 0.25 |
| Kendall's W   | 0.12 |      |      |      |       |      |

**Table 2 - Random AV Ranking**

Figure 1 shows the agreement within the group, the x-axis represents the average ranking for the set of all 6 experts, and each of the corresponding experts' ranks are shown on the y-axis. The height of the columns denotes how many experts assigned a particular ranking to an AV. Again, for comparison, Figure 2 shows the result of a group of random rankings.

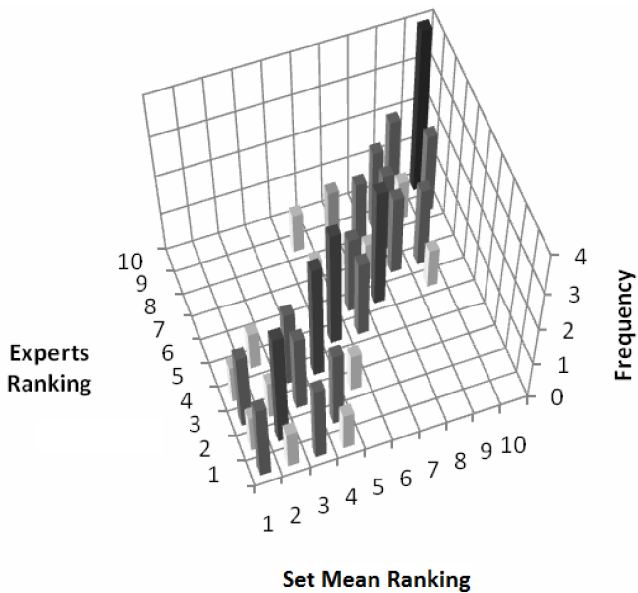

**Figure 1. Group Agreement Initial Exercise**

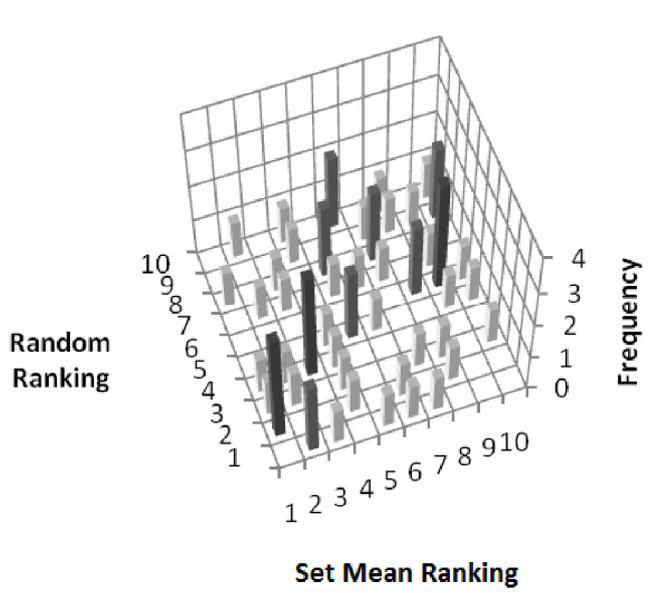

**Figure 2. Group Agreement Random**

The results of this initial exercise show that while this group of experts may have different areas of expertise, and the scenario contains omissions and uncertainties, the rankings produced by the experts have a high degree of correlation, suggesting that there is a consensus about which AVs are more difficult than others. The comparison with random rankings shows that this level of agreement is not an artefact of the averaging process. These findings were taken into consideration when creating the main exercise which includes a larger and more disparate group of experts, the aim being to explore how decision making varies within and between groups of individuals from different parts of the cyber-security community.

## Prototype Exercise

Although the previous study shows that using the experts' AV rankings we can demonstrate that there is a consensus of opinion, when it comes to the hop data we are unable to perform a detailed analysis. This is because for each hop, experts have only three possible answers, making any analysis lack the level of detail required to produce a meaningful result.

To address this, a novel approach to capturing expert opinion and expressing subjective uncertainty was devised that allows participants to make a more detailed differentiation between hops' difficulties and their certainty. There are a number of reasons that an expert could be uncertain in their answer, including:

1. The individual is not familiar with a particular technology.
2. The inherent uncertainty caused by insufficient detail in the scenario (e.g., precise component and frequency of patching not specified; in some cases this will make little difference while in others it could be significant).
3. The individual's personality
    (e.g., they may be naturally cautious about making a precise prediction).

Experts gave their answers as an interval, on a scale of 0 to 100. This was done by drawing an ellipse as shown in Figure 3, which shows an example question with two possible answers, one more uncertain than the other. The interval is produced using the points where the ellipse intersects the scale. The width of the interval denotes the uncertainty the expert has in their answer, the wider the interval, the less certain the expert is. Using this refined method, participants are able to impart much more information about their uncertainty than was possible in the initial exercise.

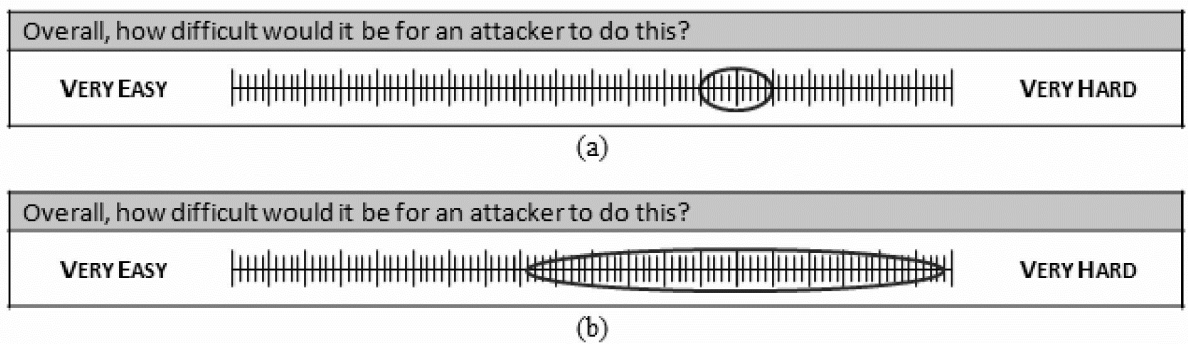

Figure 3. Interval Response where (a) is a Less Uncertain Response and (b) is a More

Uncertain Response

As a trial for the new method of eliciting expert knowledge, an exercise was carried out with PhD students and researchers from the Intelligent Modelling and Analysis research group at the University of Nottingham. Participants were asked to rate aspects of a series of restaurants in the Nottingham city centre area. For example, questions included 'How polite are the staff?' and 'Overall, how would you rate this eating place?'. Ratings were described using the proposed method of drawing ellipses (Miller 2012).

The overall response to the use of ellipses was a positive one. The participants liked the use of a single pen-stroke to determine their answer and their uncertainty.

## Main Exercise

A set of thirty nine security professionals from seven groups took part in the main exercise, drawn from a mixture of government and commercial backgrounds. The groups included system and software architects, technical security consultants, penetration testers, vulnerability researchers and specialist systems evaluators. All participants have a high level of expertise, with both breadth and depth of experience.

The reason for using a larger set of experts from multiple groups is to allow examination of the variation and agreement within each group, and between groups to see if different specialist fields of cyber-security differ in their variation and agreement, and how each field varies in its agreement with other groups.

## Data Acquisition

Following the prototype exercise, the main exercise was undertaken using the same method of eliciting expert opinion. The scenario, AVs and hops from the initial exercise were revisited and refined in order to provide a clearer definition for the participating experts. The experts were given a presentation by the scenario creator with details and diagrams of the updated scenario, AVs (see Figure 4) and hops, and had the opportunity to ask questions about the system. As in the initial exercise the experts were asked to assume that the software/hardware and frequency of patching was of the typical standard that they came across in their work with this type of government system. They all regularly work with UK government BIL3 systems, and so are aware of associated security policy and how it is typically applied to such systems in terms of component configuration, frequency of anti-virus updates, etc.

To illustrate the exercise, a diagram of one of the AVs and its constituent hops is reproduced in Figure 4. The diagram shows the system, hops and path an attacker would take to complete this attack vector. The experts were presented with this type of diagram for each of the ten attack vectors. This AV is called 'Malformed document via email', and entails an attacker on the Internet sending an email to a system user which contains a malformed PDF document. The malformed document contains a malicious exploit that compromises the desktop client and establishes a presence there in order to mount an ongoing attack. In addition to crafting the malware, the attacker must also evade detection by the relevant gateway defences, and must overcome the lockdown applied to the client. There are five distinct hops for this AV:

1. Bypass gateway content checker (i.e., evade detection)
2. Bypass gateway anti-virus
3. Compromise PDF renderer
4. Bypass anti-virus on client
5. Overcome client lockdown (i.e., access controls on the client)

A Systematic Approach to Secure System Design 11

Note that there is not a hop to bypass the gateway firewall in this attack because the exploit is carried within legitimate business traffic (i.e., an email), hence no work is required to pass through the firewall.

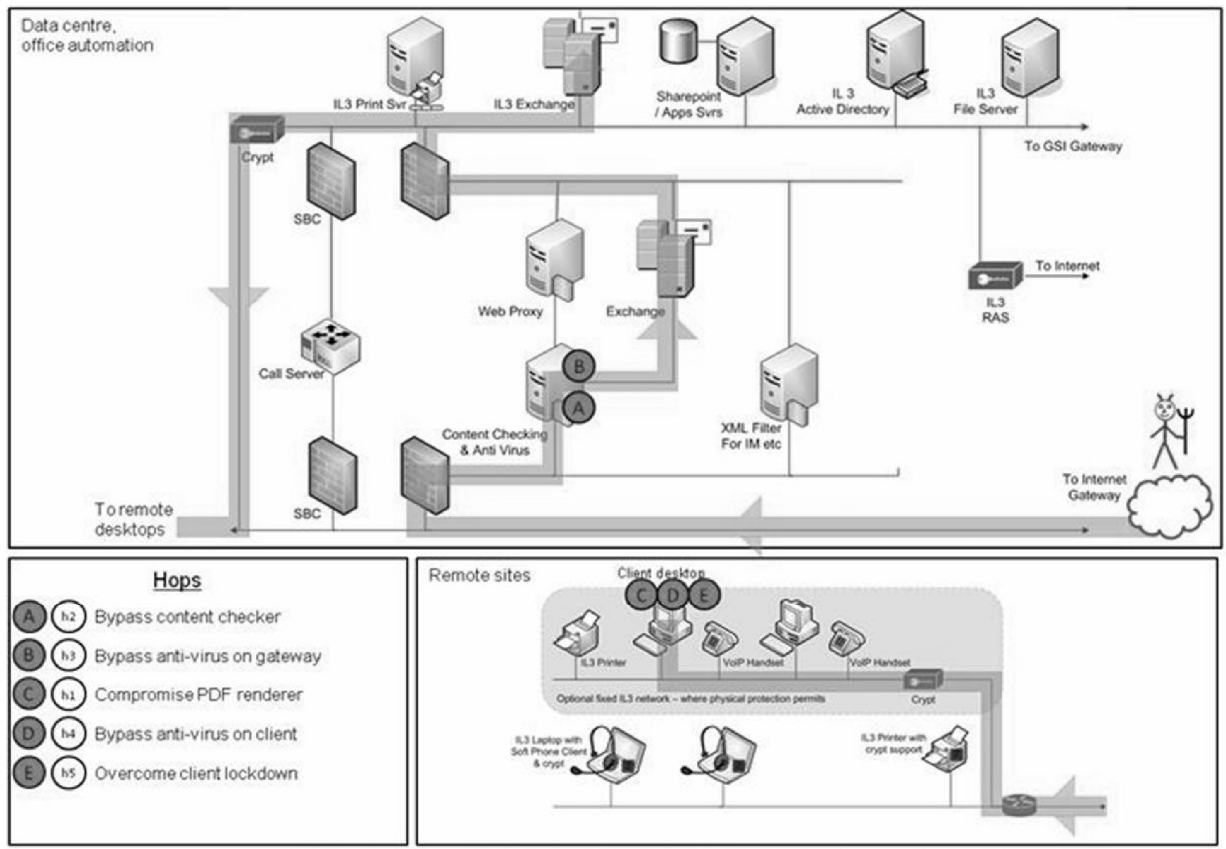

**Figure 4. Example Attack - 'Malformed document via email'**

The ten AVs that were presented to the experts are as follows:

- Malformed document via email (see Figure 4)
- Compromise central cryptographic device
- Attack via VOIP client
- Attack via network management tools
- Steal credentials and upload malicious document
- Attack via enterprise services
- Entice user to malicious website
- Subvert SAN via virtualisation infrastructure
- Instant Messaging client
- Malicious SQL attack

Note that these are shown in random order; this is because detail of how specific AVs were ranked has not been released as part of this dataset. The first part of the exercise consisted of the experts ranking the ten AVs in order of how difficult they thought it would be to carry out the attack without being detected. This activity was conducted in examination-like conditions to ensure that there were no outside influences on the experts' opinions.

Following this, the second part of the exercise required experts to answer a series of questions about each of the hops making up the AVs, giving their answers using an ellipse as in the prototype exercise. The questions were devised in collaboration with CESG's technical experts to determine what the important factors are that contribute to the difficulty of hops and AVs. The resulting questions included (but were not limited to):

1. How mature is this type of technology? (i.e., the component's technology)
2. How likely is it that there would be a publicly available tool that could help with this attack?
3. How much does the target component process/interact with any of its data inputs?
4. How complex is the target component (e.g. in terms of size of code, number of subcomponents)?

The experts were divided up, completing the hop questions in a number of separate rooms in exam conditions. As there were thirty nine participants and twenty six distinct hops in the scenario, with up to eight questions per hop, this produced a substantial dataset of around six thousand observations. We believe this scope of data collection and quantity of data collected from highly experienced security practitioners to be unprecedented. This larger number of participants and groups allows us to look at the variation between the individuals within groups, between groups, and in the overall opinions of the set of thirty nine experts.

### Data Analysis – AVs

In the first stage of analysis, work focused on the AV rankings, assessing the level of variation that occurred between individual experts, and groups of experts. Initially, individual experts' opinions were studied. Figure 5 shows how the set of thirty nine experts ranked one of the AVs, AV1. Each point on the *x*-axis represents one of the thirty nine experts, and the *y*-axis is used to show how they ranked AV1.

Two things are apparent from Figure 5:
1. There is a general consensus that AV1 is easy compared to the other AVs (1: Easiest, 10: Hardest). It can be seen that the majority of participants have given a higher (easier) ranking to AV1.
2. There is a very broad spectrum of opinions. There is at least one individual in the set that has given each of the complete range of rankings 1 – 10.

A Systematic Approach to Secure System Design 13

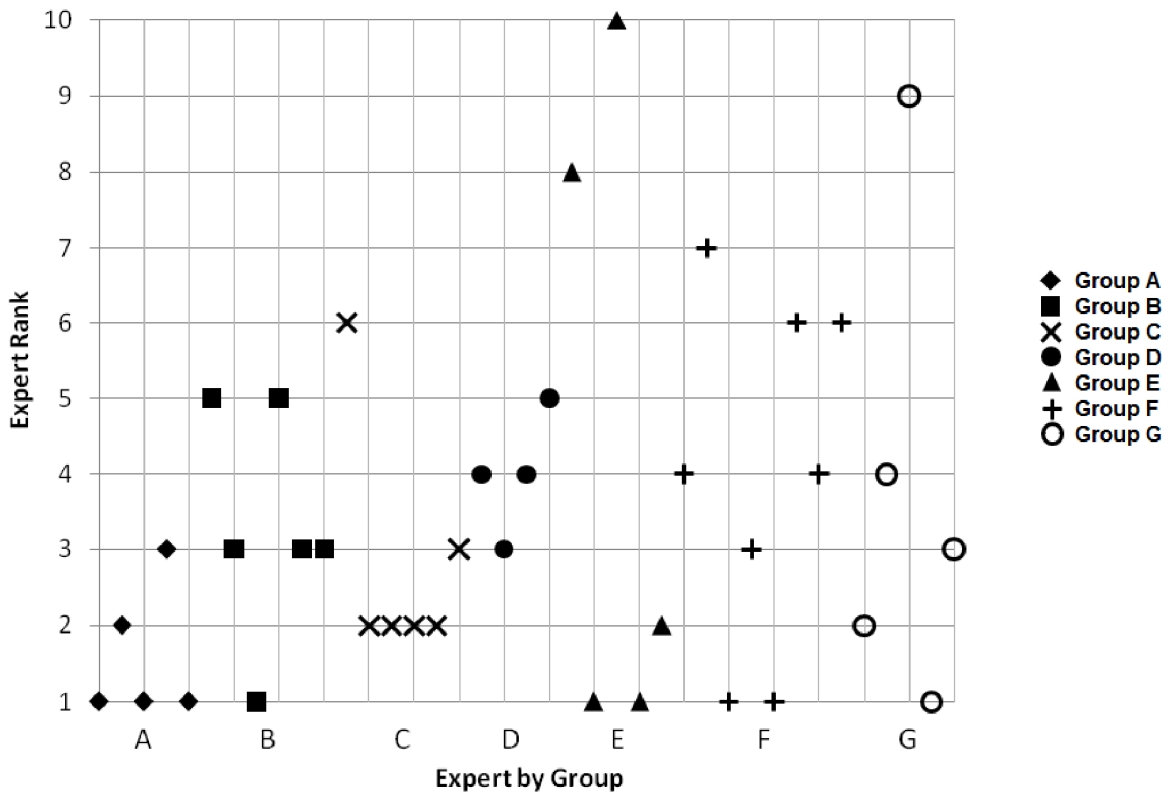

Figure 5. Spread of Rankings for AV1

The other nine AV rankings show a similar level of variation to the one shown in Figure 5, with nine of them being ranked in at least eight of the ten possible places. This tells us that while there is agreement within the set, if we were to ask one single expert for their ranking for AV1, we might get any one of the ten possible rankings. From a security advisory perspective this is obviously undesirable, as our aim is to provide clear and consistent advice. However, as we will see, by aggregating the experts' opinions we can produce a more consistent response.

At group level there is a range of levels of agreement, some of the seven groups are more consistent than others. Table 3 details the mean Spearman's rho for individuals of each group (compared against the group average) and the Kendall's W for each group. It can be seen that Group D has the best agreement, and Group G has the least agreement among its members.

| Group | A | B | C | D | E | F | G |
|---|---|---|---|---|---|---|---|
| **Mean Spearman's** | 0.82 | 0.76 | 0.83 | 0.92 | 0.61 | 0.66 | 0.56 |
| **Kendall's W** | 0.69 | 0.59 | 0.72 | 0.88 | 0.39 | 0.47 | 0.33 |

A Systematic Approach to Secure System Design 14

Figures 6 and 7 show agreement among the individuals of Group D and Group G respectively. As in Figures 1 and 2, the *x*-axis represents the group average ranking, and the *y*-axis is used to show how each expert ranked each of the AVs from the group ranking. The height of the columns denotes how many experts assigned each ranking to an AV. In these two cases there are ties in the rankings, meaning that not all of the group ranks are integer values. Obviously, as each individual has a contribution to the group mean we would expect to see some agreement, but it is clear that Group D are more consistent than Group G in their decision making.

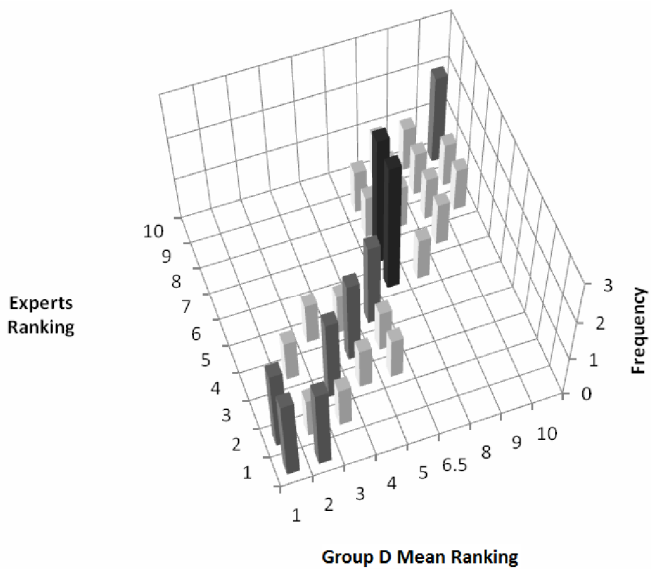
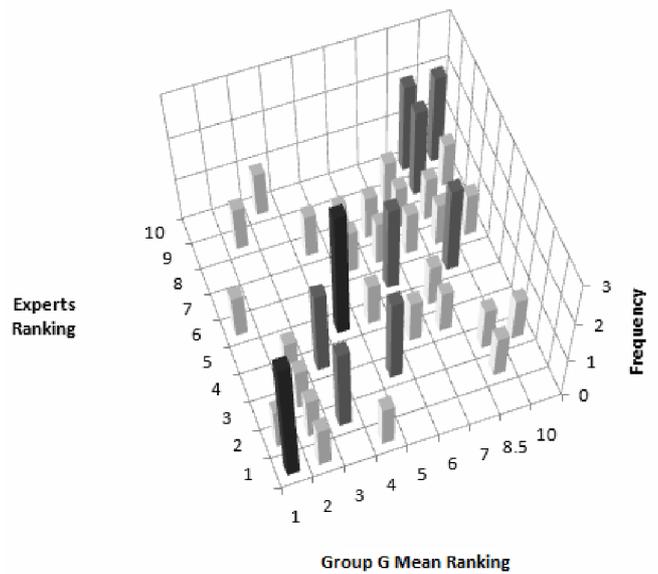

**Figure 6. Group D Agreement**  **Figure 7. Group G Agreement**

The next step in the analysis was to look at the entire set of thirty nine experts. Table 4 provides the Spearman's rho for each individual's ranking compared with the set's mean ranking. The majority of the set have a strong correlation with the overall consensus opinion, and five individuals have a particularly weak correlation with the group. By identifying outliers, we provide the opportunity to investigate further to see why these individuals disagree with the group, and whether further action is required e.g., further training, or dissemination of new knowledge.

| Group Expert | A | B | C | D | E | F | G |
|---|---|---|---|---|---|---|---|
| .a | 0.9152 | 0.6848 | 0.7212 | 0.9152 | 0.2121 | 0.9394 | 0.9273 |
| .b | 0.7091 | 0.7212 | 0.8182 | 0.9152 | 0.7576 | 0.0545 | 0.6727 |
| .c | 0.8303 | 0.5879 | 0.8667 | 0.9273 | -0.0182 | 0.8424 | -0.1636 |
| .d | 0.7697 | 0.7333 | 0.8545 | 0.7818 | 0.8545 | 0.7939 | 0.4788 |
| .e | 0.2848 | 0.6848 | 0.8667 | | 0.8788 | 0.7697 | 0.7818 |
| .f | | 0.6606 | 0.7212 | | | 0.3212 | |
| .g | | | | | | 0.9636 | |
| .h | | | | | | 0.4909 | |
| | Denotes particularly strong correlation with the group (Rho>0.7) | | | | Denotes particularly weak correlation with the group (Rho<0.3) | | |

**Table 4 - Individuals vs. Overall Group**

Table 5 and Figure 8 show each group compared with the overall set of thirty nine experts' mean ranking. In Figure 8 the *x*-axis shows the AVs in order of average rank, and the *y*-axis shows how the experts ranked each AV. It can be seen that although at individual level there is some disagreement, when the opinions of each group are aggregated there is a very strong consensus of opinion providing a consistent measure of AV difficulty.

| Group | A | B | C | D | E | F | G |
|---|---|---|---|---|---|---|---|
| **Rho** | 0.891 | 0.912 | 0.927 | 0.948 | 0.827 | 0.954 | 0.948 |

**Table 5 – Groups Compared with Set Mean Ranking**

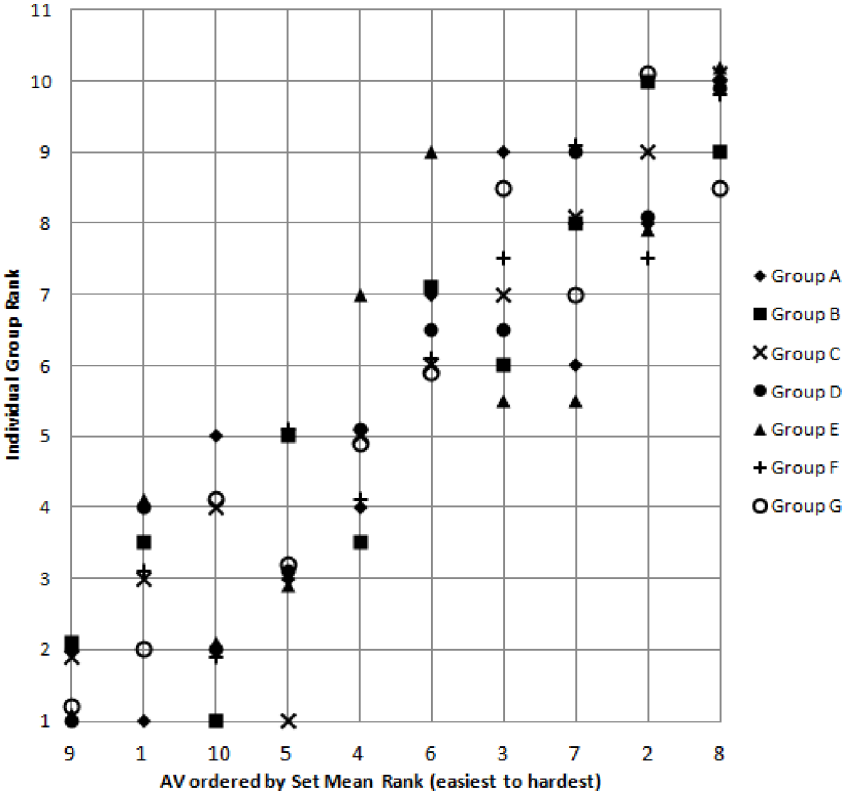

**Figure 8. All Groups – Inter-group Comparison**

Following this, scatter plots were produced to illustrate how the individuals within each group related to one another, and how the groups related to each other. Two distances were used for the plot, distance from the set mean ranking, and distance from the scenario creator's ranking. The scenario creator is the most senior member of the internal technical teams, and as the creator we can assume that he has the clearest understanding of the scenario. Because of this, his ranking can be used as a reference against which to rate others' responses (although this is not necessarily considered as 'correct'). Figure 9 contains the resulting scatter plot.

**Table 3 - Group Agreement**

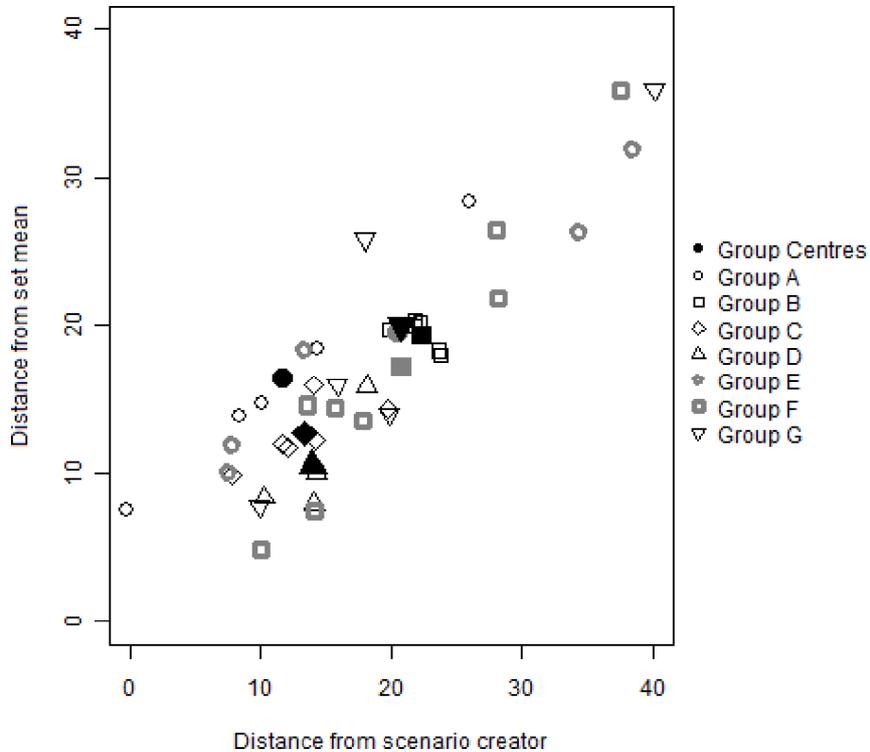

**Figure 9. Scatter plot showing individuals' distance from group and scenario creator**

As we might expect some of the groups appear to be more disparate than others. For example Group D is closely grouped, while Group G is spread out more widely, as reflected by the intra-group agreements shown in Figures 6 and 7. There may be many reasons for this, and it is not necessarily the case that Group G is inferior to Group D. For example, it may be that Group D are experts who work together regularly in the same office, and Group G could be individuals who work in different parts of the country and have never met, and therefore have very different experiences. However, full details of the groupings are restricted, and as such were not released as part of this data set, preventing a full examination of the reasons for disparity.

## Data Analysis − Hops

The second stage of analysis involves examining the data collected regarding the individual hops, and potentially relating these to the AVs. As stated previously, the hop rating part of the exercise required experts to answer a series of questions about the twenty six hops that make up the ten AVs that were ranked in the previous part of the exercise. The result of the exercise was a collection of intervals that describe each expert's opinion of a particular aspect of a particular hop.

For our initial hop analysis, we have focused on one question 'Overall, how difficult would it be for an attacker to [successfully make the hop]?' This question was designed to elicit one overall difficulty rating for the hop, while the other questions focus on specific aspects of that difficulty. This initial hop analysis is restricted to Group D, as they were the most consistent in the AV ranking part of the exercise.

Each of the twenty six hops belongs to one or more of the ten AVs, Table 6 shows which hops belong to each AV.

| AV | Hops |
|---|---|
| 1 | 2,3,1,4,5 |
| 2 | 6,7,6,8,4 |
| 3 | 9 |
| 4 | 10,11,4,5 |
| 5 | 12,13,2,3,14,15,4,5 |
| 6 | 16,16,17,4,5 |
| 7 | 6,18,4,5 |
| 8 | 19,20,21 |
| 9 | 22,23,24 |
| 10 | 25,26,1,4,5 |

**Table 6 - AVs with Constituent Hops**

Using this information, the ratings an expert gave for each hop can be used to compute a difficulty value for each AV. This has been done using a number of methods of aggregation:

1. Sum
2. Minimum
3. Mean
4. Maximum
5. Ordered Weighted Average

Each of these methods has been applied to the minimum, mean and maximum of the hop intervals for each AV. For example, Table 7 shows Expert D.b's responses to the 'Overall' question for each of the hops in AV1 (2,3,1,4,5).

**Table 3 - Group Agreement**

| Hop | Min | Mean | Max |
|-----|-----|------|-----|
| 1 | 10 | 25 | 40 |
| 2 | 30 | 40 | 50 |
| 3 | 11 | 20.5 | 30 |
| 4 | 30 | 40 | 50 |
| 5 | 60 | 70 | 80 |

**Table 7 - Expert D.b's 'Overall' Intervals for AV1**

Any of the aggregation operators above can be applied to these data to compute a difficulty score for AV1, e.g., mean of mean values (39.1), sum of maximum values (250) or minimum of minimum values (10). This process can then be repeated for each method for all ten AVs to produce scores for each method and a value for each AV. The difficulty scores are then used to rank the AVs, the ranking produced by each method is compared to the actual ranking given by the expert in the previous exercise. By doing this, the relationship between an expert's AV ranking and their hop ratings can be examined.

Discussions with a smaller group of CESG experts had generated a hypothesis that the difficulty of a given AV may be determined largely by the maximum difficulty of its constituent hops. However, it became apparent that the use of the maximum or minimum operators in particular produced a lot of equal scores for AVs, making the resulting AV ranking less meaningful. To help overcome this difficulty an Ordered Weighted Average (OWA) (Yager 1988) operator was selected as one of the aggregation methods. An OWA allows more weight to be given to the most difficult hops, while still taking into account the other hops when rating an AV. This results in a significant reduction in the number of ties obtained, leading to more meaningful rankings.

An OWA consists of a set of weights (that add up to 1), and a set of objects. In our case the objects are hop ratings. The first step of the OWA is to sort the objects (hop ratings) into descending order, so the most difficult hop will be placed at the start of the list. Then, each of the weights is multiplied by the corresponding object, so the first weight is multiplied by the first object and so on. If the first weight is high (near to one), then the resultant operator is close to a maximum. This weighting will then be reflected in the overall score produced for an AV. A selection of OWA operators were used that gave precedence to hops with higher difficulty ratings.

As an added complexity, each AV has a different number of hops, so fixed weights cannot be used with the OWA. Two alternative weighting strategies were used for the experiment, referred to as OWA(1) and OWA(2). Both use a form of ranking proportionate weighting, OWA (1) features a steady decrease of weightings, for example an AV with five hops is given a linearly decreasing weighting {5/15,4/15,3/15,2/15,1/15}. OWA (2) features a much sharper decrease in weights after the most difficult hop, for example an AV with five hops uses weights in proportion {1/2,1/4,1/8,1/16,1/32}, with the weights normalised to ensure that they sum to one.

Each aggregation operator was applied to the hop rankings given by each of the members of Group D. Then, for each method, the ranking produced was compared with the actual ranking given by the individual expert. Table 8 gives a summary of the best results found using the aggregation operators. The figures shown are Spearman's rho for the comparison between the rank derived from the hop data using each of the aggregation methods, and the actual rank provided by that individual.

| Expert | D.a | D.b | D.c | D.d | Mean |
|---|---|---|---|---|---|
| **Mean Mean** | 0.758 | 0.879 | 0.830 | 0.879 | 0.836 |
| **OWA Mean (1)** | 0.830 | 0.879 | 0.867 | 0.879 | 0.864 |
| **OWA Mean (2)** | 0.830 | 0.855 | 0.891 | 0.879 | 0.864 |

**Table 8 - Summary of Hop Aggregation Results**

The Spearman's rho figures show that there is a very high level of correlation between the rankings that the experts provided in the first part of the exercise, and those derived from their hop ratings. That is, by combining an individual's hop ratings it is possible to produce AV rankings that are closely correlated to that same individual's AV rankings. The benefit of using OWA operators that give more weight the most difficult hops can be seen, as they avoid the tied rankings produced when using a maximum operator.

## Discussion

In the methodology presented in this paper, the subjective opinions of cybersecurity professionals are used to create a consistent validation of proposed software systems. As well as providing a metric of security, it also allows systematic identification of:
1. Topics where there is a clearly established consensus of opinion.
2. Topics where there is significant disagreement between experts.
3. Individuals within the community who are consistently making judgements which are strongly away from the norm.

Note that the consensus opinion is not necessarily presented as the 'ground truth'; however, if the level of expertise is high and the degree of agreement is strong then it is more likely to hold true. In the real world, the use of measured expert opinion is often the only way of achieving a practical, realistic assessment. We also emphasise that individuals who are making 'outlier decisions' should not be considered as 'wrong'. In fact, in some cases they may represent people at the forefront of new knowledge creation. In these cases the individuals have an important role to play in challenging the 'group think'. However, in other cases, 'outliers' may simply be less experienced or skilled than the norm. If this is the case, then these individuals could be considered for targeted training. The identification of, and subsequent interaction with 'outliers' therefore needs to be carefully managed.

**Table 3 - Group Agreement**

In this research we have used the expertise of cybersecurity experts to determine a set of AVs to work with, we suggest this is the most practical method, as identifying meaningful AVs automatically is currently a very difficult task for computers to undertake. We've used what the experts believe to be the ten most salient AVs, though the optimal number to use will undoubtedly vary from system to system dependent upon its complexity. A system is only as secure as its weakest element, so identifying which are the weakest aspects of a system can form an important component of system security assessment, though obviously it is not the complete solution. Further experimentation is required to ascertain the most suitable number of AVs to use for a particular system.

While only an initial study, the analysis of the hop data has produced some interesting results. Using a limited set of data, it has been demonstrated that we are able to aggregate hop ratings for an individual to produce a measure of security for an AV. The rankings produced using these measures are closely correlated with the experts' actual AV rankings. From this we can gather that for these experts, their answer to the 'Overall' question for each hop is strongly related to how they rank AVs containing this hop. While this may seem obvious, it is by no means guaranteed a priori, due to the fact that the AV ranking and hop analysis were carried out at different times of the day, and the mapping of hops to AVs is non-obvious.

If this is the case for all of our experts, it suggests that ratings of hops can be used to rank AVs in order of difficulty. In the future, it may be possible to build a database of hop ratings that can be used to rank AVs in proposed systems, highlighting those that present greater risk.

Possible applications of the work presented in this paper include:
1. A validation tool, providing guidance on the difficulty of AVs for a proposed software system, based upon experts' knowledge of both hops and AVs. This would be very useful when designing information systems, as it could provide validation of security for given parts of a proposed system.
2. A methodology for establishing a consensus opinion of multiple experts. The aggregation process has been shown to reduce the effects of outliers, producing a more consistent measure of security.
3. A system for identifying outlying experts who may have cutting edge knowledge, or require support and training. In either case, the ability to identify outlying experts is critical to our goal of providing clear and consistent advice to system designers.

**Conclusions and Future Work**
In this paper we show how expert opinion of security can be elicited, measured and aggregated to produce an assessment for a proposed software system. The inherent variation that occurs with human experts is reduced, producing a consistent outcome. We present analysis and results of an exercise involving a set of thirty nine security professionals from seven groups including government and commercial groups, who were given a scenario created by CESG security professionals and asked to provide their opinions on ten AVs, and the twenty six hops that made up the AVs. We believe that a study of this scope and scale with highly experienced security practitioners to be unique. The analysis of the AV data showed that at individual level there was variation between individual experts. However, further analysis showed that by aggregating rankings we are able to produce rankings that are more consistent, indicating that there is overall consensus.

The initial analysis performed on hop data produced results that demonstrate how an individual's hop rankings can be used to produce a measure of security for an AV, and that the resulting rankings are highly correlated with individuals' actual AV rankings. This result tells us that there is a clear relationship between the difficulty of AVs and the difficulty of constituent hops.

Future lines of research are to include further study of the data collected during the hop rating exercise. Specifically, this will include widening the scope of the initial analysis to include all experts to determine whether the results of this study hold for less consistent groups. We will also explore the answers to the remaining questions asked about each hop in relation to the 'Overall' question, and the AV rankings. Other areas of interest with regard to the hop data include analysis of the uncertainties expressed by the experts, and determining consistencies between experts. More generally, real-world applications of the work will be considered, including providing security validation for system designers, establishing a consensus among experts, and identifying outliers.

## Acknowledgements

This work was funded by CESG - the UK Government's National Technical Authority for Information Assurance (IA).